# Non-rigid band shift and non-monotonic electronic structure changes upon doping in the normal state of the pnictide high-temperature superconductor $Ba(Fe_{1-x}Co_x)_2As_2$.

Paolo Vilmercati[1,2,#], Sung-Kwan Mo[3], Alexei Fedorov[3], Michael McGuire[4], Athena Sefat[4], Brian Sales[4], David Mandrus[2,4,5], David J. Singh[6], Wei Ku[7,8,!], Steve Johnston[1,2], Norman Mannella[1,2,*]

1. *Department of Physics and Astronomy, University of Tennessee Knoxville, 1408 Circle Drive, Knoxville, TN 37996, USA*
2. *Joint Institute for Advanced Materials, University of Tennessee Knoxville, 425 Dougherty Engineering Building, Knoxville, TN 37996, USA*
3. *Advanced Light Source, Lawrence Berkeley National Laboratory, Berkeley, California 94720, USA*
4. *Materials Science and Technology Division, Oak Ridge National Laboratory, Oak Ridge, Tennessee 37831, USA*
5. *Department of Material Science and Engineering, University of Tennessee Knoxville, 1512 Middle Drive, Knoxville, Tennessee 37996, USA*
6. *Department of Physics and Astronomy, University of Missouri, Columbia MO 65211-7010*
7. *Condensed Matter Physics and Materials Science Department, Brookhaven National Laboratory, Upton, New York 11973, USA*
8. *Physics Department, State University of New York, Stony Brook, New York 11790, USA*

*#* *pvilmer1@utk.edu*
\* [nmannell@utk.edu](nmannell@utk.edu)
*!* Present address: Department of Physics and Astronomy, Shanghai Jiao Tong University, Shanghai 200240, China

**ABSTRACT**

We report systematic Angle Resolved Photoemission (ARPES) experiments using different photon polarizations and experimental geometries and find that the doping evolution of the normal state of Ba(Fe$_{1-x}$Co$_x$)$_2$As$_2$ deviates significantly from the predictions of a rigid band model. The data reveal a non-monotonic dependence upon doping of key quantities such as band filling, bandwidth of the electron pocket, and quasiparticle coherence. Our analysis suggests that the observed phenomenology and the inapplicability of the rigid band model in Co-doped Ba122 are due to electronic correlations, and not to either the size of the impurity potential, or self-energy effects due to impurity scattering. Our findings indicate that the effects of doping in pnictides are much more complicated than currently believed. More generally, they indicate that a deep understanding of the evolution of the electronic properties of the normal state, which requires an understanding of the doping process, remains elusive even for the 122 iron-pnictides, which are viewed as the least correlated of the high-$T_C$ unconventional superconductors.

**INTRODUCTION**

High temperature (high-$T_C$) superconductivity emerges from a so-called normal state as the temperature is lowered. In conventional superconductors, a clear distinction occurs between the Landau quasiparticles forming the Cooper pairs and the phonons that are responsible for binding them [1]. This same distinction does not occur in unconventional superconductors. Here, the possible relevance of electronic and spin degrees of freedom to pairing raises the question of how to correctly describe a situation where the paired quasiparticles are also playing a role as the pairing mediators. Answering this question requires a deep understanding of the normal state, including the quasiparticles and their excitation spectrum. Obtaining such an understanding, however, has been a notoriously challenging task due to varying degrees of electronic correlations and competing orders in many systems. This is the case even for the iron-pnictides, which are in many cases viewed as the least correlated of the high-$T_C$ unconventional superconductors [2,3,4].

Obtaining an optimal normal state that ultimately gives rise to unconventional superconductivity typically involves introducing carriers into a host material (i.e. parent compound) via chemical substitution or insertions, a process commonly referred to as doping. In the iron-based superconductors, this process controls the balance of magnetism, nematicity, and superconductivity across the phase diagram. A deep understanding of the evolution of the normal state thus requires an understanding of the doping process. Nonetheless, the impact of various dopants on the electronic structure of the parent compounds and the validity of rigid band shift models in the iron-based superconductors remains the subject of considerable debate. This is best exemplified by the controversy centered on the role of transition metal (TM) substitutions in electron-doping the prototypical 122 pnictide $BaFe_2As_2$ (Ba122) [5].

Generally speaking, the strength of the impurity potential increases as the TM dopant is varied from Co, to Ni, to Cu [6,7]; early theoretical studies, however, found that the excess charge introduced by doping was concentrated around the TM substitution site even in the case of Co [6,8]. This was confirmed by experimental studies [9,10], leading some to conclude that TM dopants did not add additional carriers to the host bands, and instead modified the electronic properties of the system through their role as scattering centers. This conclusion, however, was at odds with angle-resolved photoemission spectroscopy (ARPES) studies that observed Fermi surfaces (FS) of Co-doped Ba122 that evolved as a function of doping in a way consistent with the introduction of additional carriers and Luttinger's theorem [11,12,13,14,15], and is also inconsistent with first principles calculations for Co doping [16]. Berlijn et al. [17], provided further insights into the problem by performing super-cell calculations for disordered Co substitutions. They found that Co doping lead to an electronic structure that was significantly smeared by disorder, but with FS volumes consistent with the Luttinger's theorem. Their calculations also revealed that the weak potential of the Co atoms did not produce a localized impurity state, indicating that the excess charge was indeed donated to the host bands; however, a significant amount of their spectral weight was transferred to incoherent states appearing at higher binding energies.

These results were later validated by a direct comparison of ARPES data in TM-doped Ba122, i.e. $Ba(Fe_{1-x}TM_x)_2As_2$, TM = Co, Ni, Cu [5]. There it was found that most of the doped electrons participated in the formation of the FSs, but that Luttinger's theorem was not satisfied for Ni-and Cu-doped Ba122. Specifically, the number of doped electrons ($n_{el} - n_h$) in Ni- and Cu-doped Ba122 (estimated as the difference between the electron- and hole-FS volumes) was found to be smaller than the nominal number of extra electrons (i.e. 1x for Co, 2x for Ni, and 3x for Cu), especially for Cu. This discrepancy increased from Ni to Cu, consistent with the notion that the impurity potential of the substituted TM atoms becomes stronger in going from Co, to Ni, to Cu. From a chemical point of view, Fe is divalent, and so donation of charge to the valence bands from the substitution with Co, Ni, and Cu implies chemically reasonable trivalent Co, and plausible tetravalent Ni, but would require a highly disfavored and unlikely pentavalent state for Cu. Therefore, it is not surprising the Cu cannot dope Ba122, while both Co and Ni doping have been shown to yield similar superconducting behavior when normalized for the expected number of carriers, 1x and 2x, for these elements. This also suggests that high levels of Ni doping may lead to formation of compensating defects and sample dependence related to preparation conductions especially for high Ni contents. In any case, ($n_{el} - n_h$) for Co-doped Ba122 agreed with the nominal electron count x > 0.07, where the antiferromagnetic order disappears.

These results indicate that Co-doping leads to changes in the FS consistent with Luttinger's theorem. This has helped shape the belief that the electronic structure of Co-doped Ba122 evolves according to the prediction of the rigid band model, while Cu- and Ni-doping deviates from this picture. If true, this would make Co-doped Ba122 a prototypical system to study, as the effects of doping would be well understood. However, it is worth noting that the super-cell calculations by Berlijn et al. showed a transfer of spectral weight from the coherent quasiparticles into incoherent states such that the Luttinger's count does not correspond to the total number of coherent quasiparticles [17]. This suggests that the doping evolution of the electronic properties in Co-doped Ba122 may be indeed non-trivial, in particular in regards to the band structure evolution, influenced by the interplay of disorder, correlation effects, and spin-fluctuations.

Motivated by this, we report a systematic ARPES study of Ba(Fe$_{1-x}$Co$_x$)$_2$As$_2$ for Co concentrations x = 0 to 0.22 in order to study the doping evolution of its electronic structure in proximity of the Fermi level ($E_F$). Here, the systematic use of different photon polarizations and experimental geometries allows us to disentangle the contributions of the bands forming the FS, along with their orbital symmetry. The data reveal that the normal state of Ba(Fe$_{1-x}$Co$_x$)$_2$As$_2$ and its doping evolution is indeed anomalous and inconsistent with the predictions of a rigid band model. The band filling and quasiparticle coherence are highly non-monotonic against Co concentrations. These findings demonstrate the complexity of the normal state in the pnictides, challenging current descriptions based on state-of-the-art theoretical approaches.

**METHODS**
**Samples -** High quality single crystals of Ba(Fe$_{1-x}$Co$_x$)$_2$As$_2$ were grown out of a mixture of Ba, CoAs, and FeAs, with excess CoAs and FeAs used as a flux. For example, optimally-doped crystals were grown using Ba:FeAs:Co:As=1:4.45:0.55. Each of these mixtures was heated for ~20 h at 1180 °C, and then cooled at a rate of 1 to 2 °C /h, followed by a decanting of the flux at 1090 °C. The crystals had plate-like morphologies and dimensions up to about $8 \times 5 \times 0.1$ mm$^3$, thinnest along the c-direction. A Hitachi S3400 scanning electron microscope operating at 20kV was used for determining chemical compositions using energy dispersive spectroscopy (EDS).

Multiple spot measurements and/or 250 micron line scans were taken from surfaces of as grown crystals or from surfaces exposed after cleaving to remove surface flux contamination. No significant inhomogeneity of the Co concentration through the thickness of the crystals is expected. This is based on comparison of composition measurements from as grown and cleaved surfaces, and from the sharpness of the phase transitions observed in crystals produced from this growth technique [16]. The Co concentrations, determined by Energy Dispersive Spectroscopy, were: $x = 0\%$, parent compound (P), $x = 6\%$ underdoped (UD, $T_C = 15$ K), $x = 8\%$ optimally doped (OpD, $T_C = 22$ K), $x = 12\%$ overdoped (OD, $T_C = 17$ K), and $x = 22\%$ heavy overdoped (HOD), for which no superconducting temperature $T_C$ was detected above 3 K.

**ARPES Measurements -** All of the ARPES measurements were carried out at BL 10.0.1 of the Advanced Light Source. The crystals were cleaved *in situ* in a pressure better than $3\times10^{-11}$ Torr at T = 28 K. All of the data were collected in these pressure and temperature conditions. No ageing effects were detected within a maximum of 24 hours of samples exposure to the photon beam. The total instrumental energy resolution was 18 meV, while the angular resolution was set to 0.5°, which corresponds to a momentum resolution of $\approx 0.05$ Å$^{-1}$.

The data were collected along the ΓX and ΓM directions, both in π- and σ-geometry, where the photon polarization is parallel and perpendicular to the scattering plane, respectively. Here, the scattering plane is the plane defined by the direction of the incoming photons and the direction of outgoing photoelectrons. In our experiment, the scattering plane is also a mirror plane of the sample, which is oriented perpendicular to the sample's *ab*-plane (cf. Fig. 1a).

Throughout we describe the data taken in the normal state in the folded two-Fe BZ, with the electron pocket located at the M point [$k_M = (\pi/a, \pi/a)$, Fig. 1b]. The spectra shown hereafter were excited with 60 eV photons, which probe the electronic structure at the border of the BZ along the vertical $k_z$ direction (the Z point). This choice was motivated by the fact that the FS flares out in proximity of the Z point ($2\pi/c$) [18]. As such, the separation of the Fermi crossings is maximized, making it easier to separate the contributions of different bands at $E_F$ [18]. Although the data and the band structure calculations are pertinent to the plane containing the Z point, hereafter we refer to the high symmetry directions Za and Zb as ΓX and ΓM, respectively. Note that the points denoted as "a" and "b" are not symmetry points of the three dimensional I4/mmm (139) space group.

Measurements repeated on multiple crystals of the same Co concentration were found to be completely reproducible. Since the samples used for the ARPES experiments are squares of about 1x1 mm$^2$, different spots on the surface were measured immediately after each cleave by moving the sample relative to the photon beam, whose spot is less than 100 micron in diameter. All of the measurements produced similar results, indicating good reproducibility of our results. This rules out the possibility that the observed effects are due to unequal Ba distribution after cleaving.

**Density functional theory (DFT) calculations -** The band structure calculations were carried out with the PBE-GGA functional in the virtual crystal approximation scheme. For the parent compound BaFe$_2$A$_2$ the experimental crystal structure at 175 K was used, as described by Q, Huang et al. [19]. We emphasize that we used the experimental As position.

**RESULTS and DISCUSSION**

Ba(Fe$_{1-x}$Co$_x$)$_2$As$_2$, like all pnictides, is a multi-orbital system, with FS originating from several bands comprised of the Fe 3d orbitals [3,4]. The majority orbital character of each band can be identified using the symmetry of the photoelectron's initial state, which is accessed using different photon polarizations and experimental geometries [20,21,22]. This methodology exploits the fact that the intensity in ARPES experiments is modulated by a dipole matrix element. As such, a photoelectron can be detected only if the triple product of the dipole operator, the initial state photoelectron wavefunction, and the final state photoelectron wavefunction is of even parity [20,21]. The final state of the photoelectron is of even parity when the detector slit is contained in the mirror plane, while the dipole operator is even (odd) for π- (σ-) polarization. Consequently, σ- and π-polarizations are sensitive to initial state photoelectron wavefunctions of odd and even parity, respectively. Based on these considerations, the orbital character of the initial state wavefunctions that is detected with different photon polarizations and scanning directions used in our experiment are summarized in Table 1.

**Fermi Surfaces and image plots -** The evolution of the FS as a function of Co concentration, measured along the ΓM direction for σ- and π-polarizations, is shown in Fig. 2. The M point is contained in the mirror plane for this experimental geometry. Since almost no spectral weight is observed at the M point for the maps collected in π-polarization, according to Table 1 the data indicate that the electron pocket at M derives from bands of $d_{xy}$, and/or $d_{yz}$ orbital character, in agreement with previous studies [22,23]. The decorations of the electron pocket located at (0.75 Å$^{-1}$, 0) and (1.45 Å$^{-1}$, 0) disappear during the early stages of doping. This indicates that these features are likely due to the band folding associated with the orthorhombic distortions in the P and UD compounds.

According to the rigid band model, the effect of electron- and hole-doping would be completely accounted for by simply raising or lowering the chemical potential, respectively, by an amount proportional to the introduced charge. Qualitatively, the FS maps already reveal that the dependence of the size of the hole and electron pockets on Co concentration is more complicated than that predicted by the rigid band model. In σ-polarization the size of the hole pocket shrinks progressively with increasing Co concentration. Conversely, in π-polarization no obvious reduction in size of the hole pockets occurs with the exception of the HOD sample. The size of the electron pocket, visible only in σ-polarization, does not seem to increase significantly, again with the exception of the HOD sample. This behavior is clearly at odds with the prediction of a rigid band model, which predicts monotonic changes in the FS. This conclusion is corroborated by more detailed data taken along the ΓX and ΓM directions.

The image plots of the bands at Γ and M along the ΓX and ΓM directions are shown in Fig. 3. The continuous lines superimposed over the image plots are the results of our DFT calculations, where we have applied a band renormalization of ≈ 2 which is typically required to obtain better agreement with ARPES data in proximity to $E_F$ [24,25]. The DFT calculations predict three hole-like bands at Γ and two electron-like bands at M, each with mixed orbital symmetry. The results of the DFT calculations are shown in detail for the parent compound in APPENDIX 1, with highlighted orbital character for each band (cf. Fig. S1). The predominant character of the bands forming the hole pockets are xy, xz, yz, and $z^2$, while the predominant character of the electron pockets are xy, xz and yz.

Our DFT calculations predict that the band structure of the parent compound rigidly shifts to higher energies with increasing Co concentration and provide a direct estimate of the chemical potential (μ) shift for the P, OP, and HOD compounds. Specifically, the $E_F$ shifts amount to 25 meV, 35 meV, 56 meV and 120 meV for the UD, OP, OvD and HOD compounds, respectively. The value of μ for the UD and OvD compounds was determined by interpolating the results with a second-degree polynomial. Therefore, according to the DFT calculations, an increase of the Co concentration should monotonically decrease (increase) the size of the hole (electron) pockets. This is contrary to the observations shown in Fig. 3, where we observe non-trivial changes in the normal state electronic structure as a function of Co concentration. For instance, in both the ΓX and ΓM directions, the size of the hole pocket in σ-polarization seems to reduce progressively as the Co concentration increases (Figs. 3(a), (c)); however, in π-polarization there is no obvious size reduction (except for the heavily overdoped (HOD) sample) (Figs. 3(b), (d)). Conversely, the size of the electron pocket does not increase significantly except for the HOD compound (Figs. 3(e), (f)).

**MDC Analysis and the extraction of $k_F$** - To describe this behavior more quantitatively, the Fermi momenta $k_F$, i.e. the locations in momentum space where the bands cross $E_F$, have been determined by an analysis of the Momentum Distribution Curves (MDC) at $E_F$. The MDC curves plot the distribution of counts as a function of momentum at a fixed energy (in this case $E_F$). An important aspect of our analysis is that the MDC spectra have been extracted by integrating a window of only 3 meV (i.e. ± 1.5 meV) about $E_F$. This is necessary for correctly determining the Fermi crossing of each band in a multi-orbital system like the Fe pnictides. We illustrate this point in Fig. 4, where different integration windows have been applied to data collected in the parent compound at Γ (hole pocket) along the ΓM direction with σ-polarization. The spectra have been shifted such that the peaks on the right hand side of the plot are aligned to the same position. The positions of the peaks on the left hand side drift progressively to the left for larger integration windows, clearly indicating that the choice of windows affects the identification of the Fermi momenta $k_F$. Larger integration windows also produce a substantial broadening of the MDC spectra. In principle, this effect can alter the minimum number of peaks required to fit the MDCs.

The image plots of the bands at Γ and M along the ΓX and ΓM directions, the corresponding MDC curves at $E_F$, and their fits are shown in Figs. 5 and 6, respectively. With our choice of a ±1.5 meV integration window, the extracted MDC spectra are intrinsically broad, indicating that multiple bands cross $E_F$. The spectra were fitted with the minimum number of pairs of peaks, with each pair fixed to be equidistant from the center of the spectrum. For the hole pocket, the spectra collected along both the ΓX (Figs. 5 a-d) and ΓM (Figs. 6 a-d) directions, in both polarizations, were always fit with two pairs of peaks. The first pair closer to the Γ point describes the same band since the peak positions corresponding to $k_F$ do not change when the polarization is switched. This is in contrast to the positions of the second pair, which have different positions in different polarizations, indicating that they arise from different bands. Along the ΓX direction (cf. Fig. 5 a-d) three bands cross $E_F$: one detected only in σ-polarization (HBσ$_{ΓX}$), one detected only in π-polarization (HBπ$_{ΓX}$), and an inner one detected in both polarizations (HB$_{\Gamma X}$). The situation is identical along the ΓM direction (cf. Figs. 6 a-d), with the bands denoted as HBσ$_{\Gamma M}$, HBπ$_{\Gamma M}$, and HB$_{\Gamma M}$.

The phenomenology described above is common for all of the compounds, with the exception of the parent compound. In this case, in both polarizations along the ΓM direction (cf. Figs. 6 b,c), most of the intensity is accounted for by a single peak. We assign this peak to the HBπ$_{\Gamma M}$ band, while the HB$_{\Gamma M}$ band remains undetected. Two additional peaks of relatively lower intensity are needed to fit the tails of the spectrum in the σ-polarization (see Fig. 6b). These peaks represent an additional band appearing due to the band folding induced by the orthorhombic distortion that accompanies the structural phase transition at T = 140 K [26]. Along the ΓX direction, the P compound has an additional band of relatively low intensity in σ-polarization (cf. Fig. 5b). This observation is particularly interesting since along the ΓX direction one does not expect to find any signature of the band folding. Within the error bar, this band has the same k$_F$ as the outer band detected in π-polarization, i.e. HBπ$_{\Gamma X}$, suggesting that they are the same band. For the electron pocket (Figs. 5 e,f and 6 e-f), the data are fit with two pairs of peaks, denoted as an external band (EB$_{EX}$) and an internal band (EB$_{IN}$).

To summarize our MDC analysis, we find that three bands cross E$_F$ at Γ: one band detected only in σ-polarization (HBσ$_{\Gamma X}$), one detected only in π-polarization (HBπ$_{\Gamma X}$), and an inner one detected in both polarizations (HB$_{\Gamma X}$). Similarly, two bands cross E$_F$ at M, denoted as EB$_{EX}$ (external) and EB$_{IN}$ (internal) bands. The same systematics are found for cuts along the ΓM direction. These results are summarized in Fig. 1(b).

**Identification of the orbital character -** The parity rules described in Table 1 provide guidance for identifying the orbital character of the bands at Γ and M. We note that it is not possible to assign a unique orbital character to each band unambiguously, as DFT calculations show that each band contains contributions from multiple orbitals (cf. Fig. S1). Here, our analysis identifies the majority orbital character of the band, which we refer to simply as the orbital character for brevity.

At Γ we assign a mixed $d_{xz-yz}$ character to the HB$_{\Gamma X}$ and HB$_{\Gamma M}$ bands, since they are visible in both polarizations. The HBσ$_{\Gamma X}$ has $d_{x^2-y^2}$ character, while the HBπ$_{\Gamma X}$ band has either $d_{xy}$ or $d_{z^2}$ character. (Our data suggests a dominant $d_{z^2}$ character since the photon polarization is perpendicular to the sample plane in π-geometry and the intensity of the in-plane $d_{xy}$ contribution will be suppressed.) Along the ΓM direction, bands with $d_{xz}$, $d_{x^2-y^2}$, and $d_{z^2}$ character can be detected with π-polarization, while the use of σ-polarization allows the detection of bands with $d_{xy}$ and $d_{yz}$ character. The HBσ$_{\Gamma M}$ band thus has $d_{xy}$ character. We assign a predominant $d_{z^2}$ character to the HBπ$_{\Gamma M}$ band, since in π-geometry the photon polarization is perpendicular to the sample plane and the intensity of the in-plane $d_{x^2-y^2}$ contribution is suppressed. In σ-polarization we detect an additional band of relatively low intensity along the ΓX direction of the P compound (cf. Fig. 5 b). Since this band is also visible in σ-polarization, it is possible that the HBπ$_{\Gamma X}$ band has $d_{xz-yz}$ character in the P compound, which then is suppressed when Co is inserted, suggesting that Co doping might change the mixing of the orbital characters of certain bands.

The cuts at the electron pocket along the ΓX direction (cf. Fig. 5 e-f) cannot be used to determine the symmetry of the bands, since these spectra were not collected in the mirror plane. As already discussed above, since along the ΓM direction the intensity is suppressed in the case of π-polarization (cf. Fig. 2), we assign a mixed $d_{xy}$ and $d_{yz}$ character to the $EB_{EX}$ and $EB_{IN}$ bands forming the electron pocket at M, in agreement with previous investigations [22,23]. A summary of the majority orbital character assigned to each band is provided in Table 2.

**Co concentration dependence of the Fermi momenta $k_F$: comparison with the DFT results -** The Fermi momenta $k_F$ extracted from the MDC analysis are shown and compared to the results of the DFT calculations in Fig. 7. The dependence of $k_F$ on Co concentration is clearly at odds with a rigid band shift scenario. Any attempt to match bands with the same orbital character would imply shifting each DFT band of a different amount. More importantly, the overall shift of the band structure with Co concentration is not monotonic: the size of the hole (electron) pocket does not reduce (enlarge) progressively as one would expect for a rigid shift of the bands. Only comparing the P with the HOD compound can one conclude that the insertion of Co makes the hole pockets shrink and the electron pockets enlarge, with the exception of the bands at the electron pocket along the ΓM direction. The most obvious disagreement with the DFT predictions can be observed for the bands at the electron pocket. Specifically, the position of the $k_F$ does not change along the ΓX direction from the P to the OP compound. Along the ΓM direction, $k_F$ decreases in the first stages of doping, and then increases for the OD and HOD compounds, but remains smaller than that of the P compound. Another important observation is that the dependence of $k_F$ on Co concentration changes in proximity of optimal doping: up to OpD compound there is an overall poor agreement even on the qualitative level between the data and theory, while for higher doping levels the agreement between data and theory improves on a qualitative level, but remains poor on a quantitative level. Related to this, Yi and co-workers found momentum dependent band shifts and renormalizations in this compound [27].

**Energy Distribution Curves Analysis -** Further insights into the doping evolution of the normal state electronic structure can be obtained from an analysis of the Energy Distribution Curves (EDCs) (Fig. 8). Each EDC stack consists of spectra where electron counts are plotted at fixed momenta and as a function of the binding energy (BE). The EDCs of the data collected at the Γ point (hole pocket) are shown in the sequence of Fig. 8(a-d). The four panel sequences correspond to cuts along the ΓX (a and b) and ΓM directions (c and d) with σ- and π-polarization.

The highlighted spectra denote the EDCs at the Fermi momenta ($k_F$), which are identified by the peak positions obtained from the fit of the MDCs (cf. Figs. 5 and 6). We refer to the EDCs with $k_F$'s closer to and farther from the Γ point as "inner" and "outer", respectively. The inner EDC thus corresponds to the $k_F$ of the $HB_{ΓX}$ band in panels (a) and (b), and the $HB_{ΓM}$ band in panels (c) and (d), respectively. In panels (a) through (d), the outer EDC corresponds to the $k_F$'s of the (a) $HBσ_{ΓX}$, (b) $HBπ_{ΓX}$, (c) $HBσ_{ΓM}$, and (d) $HBπ_{ΓM}$ bands. Also visible, particularly for the spectra in π-polarization, is a band at higher BE (≈ 200 meV) that does not cross $E_F$.

The EDC spectra corresponding to the image plots at the M point (electron pocket, cf. Figs. 5 (e,f) and Figs. 6 (e,f) are shown in Figs. 8 (e) and (f) for cuts taken along the ΓX and ΓM directions, respectively. The highlighted EDCs closer to and further from the M point, referred to as "internal" and "external" EDCs, respectively, correspond to the $k_F$'s of the $EB_{IN}$ and $EB_{EX}$ bands, respectively. The highlighted spectra at the center of the EDC stack denote the EDC at the M point, and correspond to the bottom of the electron pocket. In addition to the two bands crossing the Fermi level, there is a broad hole-like band that never crosses $E_F$, but intersects the $EB_{IN}$ and $EB_{EX}$ bands close to the M point. This band is more intense for cuts along the MX direction (panel e).

The EDCs at the M point, i.e. at the bottom of the electron pocket, are shown in Fig. 9 along the (a,b) MX and (c,d) ΓM directions for different Co concentrations. The raw data, shown in panels (a) and (c), again reveals that the evolution of the bands is not consistent with a rigid band shift picture. For example, the binding energy (BE) of the peaks provides an estimate of the bottom of the band at the M point. For the initial stages of Co doping, the EDC peak positions remain constant and it is only for the OD and HOD compounds that the EDC peaks shift to higher BE with increasing Co concentration, as would be expected for an increase in carrier concentration.

In order to provide a quantitative estimate of the position of the bottom of the band as a function of Co concentration, the positions of the peaks corresponding to the $EB_{IN}$ and $EB_{EX}$ bands were obtained from a fit of the EDCs. Fitting the EDC is not an easy task, since the pnictides are multi-band systems. As such, a proper analysis of the EDC spectra must take into account contributions from different bands. The results of the EDCs fit shown here are based on an iterated analysis that resulted in an overall self-consistency. Specifically, we applied a model with the minimal number of peaks, where the number of peaks was fully consistent with the results of the MDC analysis. Whenever additional peaks were needed to fit the EDC spectra at high binding energy, we verified that they accounted for bands at higher BE visible in the raw data (i.e. image plots and EDC stacks). Specifically, two peaks account for the $EB_{IN}$ and $EB_{EX}$ bands, while the spectral weight appearing at higher BE (peak denoted as B in the figure) is due to a broad hole-like band that never crosses $E_F$, but instead intersects the $EB_{IN}$ and $EB_{EX}$ bands close to the M point. This band is visible in the EDCs stack shown in Figs. 8 (e,f). The fit of the spectra of the P and UD compounds requires an additional contribution (narrow peak denoted as BF) to account for the band structure folding arising from the orthorhombic distortion.

Allowing for these contributions, the EDC spectra were fit with the most common lineshapes, namely Doniach-Šunjić (DS) and Lorentian (L), for direct comparison [28]. The DS profile (Figs. 9 (a) and (c)) is typically used to describe the lineshape of photoelectron peaks in metallic systems. This profile has an asymmetric tail at higher BE that captures the energy loss of the photoelectron's kinetic energy. As such, it accounts already for the background of inelastically scattered electrons. For the fit of the EDC with a L profile, the background was first subtracted with a Shirley routine (Figs. 9 (b), (d)).

We found that it was not possible to obtain a satisfactory fit using the DS profile (cf. Figs. 9 (a) and (c)), largely due to difficulty in accounting for the spectral weight at higher BE. For example, while the experimental data are fit well, the results are unphysical: the B peak was found to be extremely broad, spanning the whole range of the spectra, and with significant spectral weight at $E_F$. This is in contradiction of the fact that the B peak accounts for a hole-like band that never crosses $E_F$. In contrast, the best fit of the data with meaningful physical interpretation was obtained with the use of a L profile for all of the identified bands, with the addition of a broad Gaussian (G) peak to account for the spectral weight at higher BE (cf. Figs. 8 (b), (d)). We note that the use of the DS profile does not produce satisfactory results even when a broad G peak was added in order to capture the missing spectral weight at higher BE. In fact, with the addition of the broad G peak, the DS lineshape of the peaks tended to lose their asymmetry and become essentially L.

The BE of the $EB_{IN}$ and $EB_{EX}$ bands extracted from fitting the EDC spectra at the bottom of the electron pocket are shown in Figs. 10 (e,f). The dependence of the BE of the $EB_{IN}$ and $EB_{EX}$ bands on Co concentration is clearly at odds with a rigid shift of the band structure, since in this case the bottom of the bands is expected to drift progressively away from $E_F$. The Co concentration dependence of the Fermi momenta $k_F$ extracted from the MDC analysis has already been discussed in Fig. 7 when compared to the DFT results. It is, however, shown again in Figs. 10 (a)-(d). Given that the error bars are smaller than the size of the symbols, the data shown in Fig. 10 reveal a non-monotonic dependence of both $k_F$ and the BE of the bands at the bottom of the electron pocket versus Co concentration, with abrupt changes on going from UD to Opt doping. The Fermi momenta $k_F$ (cf. Figs. 10 (c)-(d)) and the BE at M (cf. Figs. 10 (e)-(f)) of the $EB_{IN}$ and $EB_{EX}$ bands are plotted together in Fig. 11, providing a schematic picture of the evolution of the electronic structure of the electron pocket as a function of Co concentration. The non-monotonic dependence of both $k_F$ and the BE of the bands at M versus Co concentration underscores that the difference with respect to the prediction of rigid band models are significant.

**Non-trivial changes in the normal state electronic structure as a function of Co concentration -** The EDC spectra at $k_F$, highlighted in Fig. 8 reveal additional non-trivial changes in the normal state electronic structure as a function of Co concentration. The EDC spectra at $k_F$ for the bands forming the hole pocket and the electron pockets are shown in Fig. 12 and Fig. 13, respectively. The evolution of the EDCs at $k_F$ with Co concentration depends strongly on the photon polarization. At the hole pocket (Fig. 12), in both $\Gamma X$ and $\Gamma M$ directions, with σ-polarization the EDCs sharpen at optimal doping, both for the outer bands (Figs. 12 (a),(c)) and for the $HB_{\Gamma X}$ and $HB_{\Gamma M}$ inner bands (cf. Fig. 12 (e),(g)). A similar sharpening is observed in the EDCs at M taken with σ-polarization, although the effect is less pronounced (cf. Fig. 13). Conversely, no sharpening of the EDCs at optimal doping is observed in π-polarization (cf. Figs. 12 (b),(d),(f),(h)). The EDC spectra are much broader in this polarization, with a larger background that appears to be maximal at optimal doping.

Since σ- and π-polarizations enhance emission from in-plane and out-of-plane initial states, respectively, the data indicate that at optimal doping the spectral features for states with in-plane character become sharper (more coherent). This phenomenology appears to be drastically different for the data around the $\Gamma$ point collected in π-polarization, which are representative of the out-of-plane character of the states.

These observations indicate that Co substitution has a dramatic effect on the states at $E_F$, which appear to be affected differently depending on their in-plane or out-of-plane character. Interestingly, the sharpening of the in-plane states is also non-monotonic, as the effect is most pronounced at optimal doping. In general, a broadening of the spectral features is consistent with a certain amount of disorder induced by Co substitution; however, this interpretation is difficult to reconcile with the observed non-monotonic behavior of the spectral width and the fact that this broadening is not present in the data collected in σ-polarization. Naively, one would expect a monotonic increase in the impurity broadening in both polarizations as the Co-concentration is increased.

The fact that the in-plane states at $E_F$ become sharper for optimal doping is certainly intriguing. A detailed mechanism for the observed phenomenology remains unclear. The data suggest a relationship between the macroscopic properties of the Ba(Fe$_{1-x}$Co$_x$)$_2$As$_2$ system and the features of our angle-resolved single particle excitation spectra with in-plane character. If the observed phenomenology is a direct manifestation of a link between superconductivity and the coherence of single particle spectra in the normal state, such link is certainly not trivial, as the coherence of the normal state quasiparticle and the coherence of the macroscopic superconducting state are completely unrelated concepts. [29,30]. Alternatively, the gain of coherence of the in-plane states could be a simple correlation to superconductivity as the latter is related to the increase of the interlayer spacing. In fact, a generic feature of Fe-based superconductors is the correlation between the superconducting transition temperature $T_C$ and the bond angles and positions of the pnictogen atom above Fe in the tetrahedral [3]. In the Ba(Fe$_{1-x}$Co$_x$)$_2$As$_2$ system $T_C$ increases approximately linearly with increasing Co content (at least until optimal doping is achieved), and the c-axis lattice constant (proportional to the interlayer spacing) decreases monotonically, while the a-axis spacing remains almost constant [31]. In light of this, the data might suggest there is a gain of coherence for the in-plane states concomitant to a decrease of the interlayer spacing. Interestingly, it has been suggested that the bonding angle of the Fe tetrahedral is correlated to the interorbital coupling strength, i.e. the degeneracy of the $d_{xy}$, $d_{xz}$, and $d_{yz}$ states [32]. It should also be kept in mind, however, that the correlations between lattice constants, bond angles, and $T_C$ are strongly dependent on the particular family of compounds [3]. Additional investigations are clearly needed in order to clarify the details of the observed phenomenology revealed by our data.

**Non-rigid band shift: significance and possible causes –** We now comment on the significance of our results for the physics of the pnictides and the possible underlying causes for the non-trivial effects revealed by our data.

We first address the possible mechanisms responsible for the deviation from a rigid band model in Co-doped Ba122. Here, it is important to first clarify the notion of rigid band shift, which entails the possibility of obtaining the band structure of the doped system by simply shifting the chemical potential for the undoped system. In general, there are several possible underlying causes for a non-rigid band shift: i) the impurity potential is not negligible, ii) the occurrence of self-energy effects due to impurity scattering, iii) manifestations of strong spin fluctuations or other electronic correlations and the ensuing renormalizations, and iv) a combination thereof. It is also possible that the observation of a non-monotonic band shifts could be a manifestation of the complexity of the phase diagram. All of the samples were measured in the normal state, at the temperature T = 28 K. Along this temperature line of the phase diagram, the P compound has an orthorhombic structure with antiferromagnetic (AFM) order, the UD sample is still very close to the AFM phase, and the OpD, OD, and HOD samples have a tetragonal structure. It is possible that the observed phenomenology could be linked to the influence of the crystallographic and magnetic phase changes on the underlying electronic structure.

For the specific case of Co-doped Ba122, the first possibility seems unlikely. Indeed, previous ARPES investigations in Ba(Fe$_{1-x}$TM$_x$)$_2$As$_2$, TM = Co, Ni, Cu support the notion that the impurity potential of substituted Co is much weaker than that of Ni and Cu [5] and that charge is indeed donated to the host bands. Self-energy effects due to impurity scattering are also difficult to understand in the context of our data. As previously stated, self-consistent T-matrix calculations, which average over disorder configurations, show that impurity scattering gives rise to self-energy that is proportional with the impurity concentration [33,34]. It is therefore difficult to reconcile the non-monotonic changes of several quantities with an increase in impurity scattering rates as a function of Co concentration. Our observations also cannot be accounted for purely from changes in the spin density wave phase. First, the location band bottom for the two electron pockets have the opposite behavior as the doping tracks from the P to the OPT doped compound, while one would naively expect both bands to shift in the same direction with the collapse of the spin density wave. Second, we observe a non-monotonic shift of the electron pocket in moving from the P to the OD compound, where no trace of the spin-density-wave is found experimentally. Finally, the magnitude of the band bottom shift, if due to a spin-density-wave, would imply a rather large reconstruction at the Fermi level in a mean-field description, which we do not observed within our experimental resolution. These factors lead us to believe that the most plausible cause of the inapplicability of the rigid band model in Co-doped Ba122 is the occurrence of electronic correlations and other many body effects, possibly due to different renormalization effects as carriers are introduced via Co substitution. This picture is consistent with the result of the super-cell calculations by Berlijn et al. [16], which indicated that the disorder potential associated with the random introduction of carriers reorganizes the spectral weight around the FS. Note that here we are using the word "correlations" as a blanket term that encompasses many body effects of possibly different nature, such as inter- and intra-orbital magnetic correlations (Hund's coupling), and Hubbard interactions. Although in the pnictides the Hubbard correlations have been found to be moderate at most, certainly they cannot be dismissed, as suggested by the renormalization of ~ 2 that is necessary to apply to the electron bands revealed by ARPES experiments in order to have a satisfactory correspondence with DFT band structure calculations [35].

The data presented so far are evidence that a rigid band shift scenario is untenable in Co-doped Ba122. In fact, the non-monotonic dependence of $k_F$, the BE of the bands at M, and the sharpening of the in-plane states versus Co concentration emphasizes that the difference with respect to the prediction of rigid band models are significant, and that the doping evolution of Co-doped Ba122 is not trivial. This invalidates the notion that Co-doped Ba122 is a prototypical system to study, as the effects of doping are much more complicated that previously believed. It is also apparent that Cu- and for some heavily Ni-doped samples, Ba122 deviate from the prediction of the rigid band model, since changes in the FS upon doping are not consistent with Luttinger's theorem [5]. The conclusion is thus inescapable: a deep understanding of the evolution of the normal state, which requires an understanding of the doping process, remains elusive even for the 122 iron-pnictides, which are viewed as the least correlated of the high-$T_C$ superconductors.

We stress that, in principle, our results are not inconsistent with previous studies that validated Luttinger's theorem in Co-doped Ba122 [5]. The issue is that the validation of Luttinger's theorem is a necessary, but not a sufficient condition for the applicability of a rigid band model.

Put differently, Luttinger's theorem must be necessarily satisfied if a rigid band model is valid, but not vice versa. In fact, we note on passing that it would be useful to extend the methodology used here (a narrow integration window about $E_F$, necessary for correctly determining the Fermi crossing of each band, and the systematic use of different photon polarizations and experimental geometries) to systematic measurements as a function of the photon energy required to sample the whole BZ along the vertical $k_z$ direction. The possibility of disentangling the contributions of the bands forming the FS, along with their orbital symmetry, for the whole BZ, as opposed to a single $k_z$ value sample as done here, will provide a more precise determination of the FS volume and the Luttinger count. This is necessary to determine with greater accuracy whether the effective carrier concentration at $E_F$ is proportional to the Co concentration, so as to validate more assertively the one-to-one correspondence between Co concentration and effective doping level at $E_F$, tacitly assumed in theoretical works.

## CONCLUSIONS

Motivated by intense debates surrounding the nature of the normal state and the validity of rigid band models in the iron-based superconductors, we have studied the doping dependence of the electronic structure in the normal state of Co-doped Ba122 in proximity of $E_F$ with ARPES. The data reveal that the normal state of $Ba(Fe_{1-x}Co_x)_2As_2$ and its doping evolution is indeed anomalous and inconsistent with the predictions of a rigid band model. Specifically, we found that any attempt to match the Fermi crossings revealed by the data with the values predicted by DFT calculations requires shifting each DFT band by a different amount. In fact, the difference with respect to the prediction of rigid band models are even more significant, as revealed by the highly non-monotonic dependence on Co concentration of the Fermi crossings, the BE of the bands at M, and the sharpening of the in-plane states at $E_F$. These behaviors in turn reflect a non-trivial dependence upon doping of key quantities such as band filling, bandwidth of the electron pocket, and quasiparticle coherence, respectively, and demonstrate the complexity of the normal state in a prototypical pnictide system.

The results of previous investigations [5], and the highly non-monotonic dependence of electronic properties on Co concentration revealed in this study seem to dismiss both the size of the impurity potential, and the occurrence of self-energy effects due to impurity scattering, as plausible mechanisms responsible for the deviation from a rigid band model in Co-doped Ba122. Although a detailed mechanism for the observed phenomenology remains unclear, a most plausible cause of the inapplicability of the rigid band model in Co-doped Ba122 seems to be the occurrence of electronic correlations, possibly due to different renormalization effects as carriers are introduced via Co substitution.

The inapplicability of a rigid band model and the anomalous doping evolution of the electronic properties in Co-doped Ba122 indicates that the effects of doping in pnictides are much more complicated that currently believed. More generally, our findings indicate that a deep understanding of the evolution of the electronic properties of the normal state, which requires an understanding of the doping process, remains elusive even for the 122 iron-pnictides.

**FIGURE, FIGURE CAPTIONS AND TABLES**

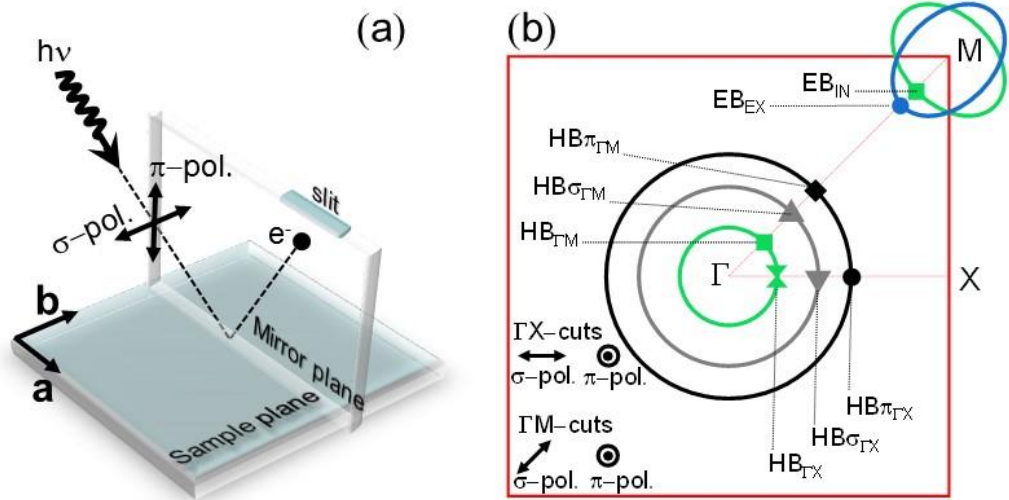

Fig. 1

**Fig. 1. Experimental Methodology.** (a) Schematic illustration of the experimental geometry. The directions of the incoming photon (hν) and the outgoing photoelectron (e⁻) define the scattering plane, which in our experiment is also a mirror plane of the sample, which is oriented perpendicular to the sample's *ab*-plane. The analyzer slit, parallel to the sample plane, is contained in the scattering plane. The photon polarization is either parallel (π) or perpendicular (σ) to the scattering plane or, analogously, parallel (σ) or perpendicular (π) to the *ab* sample plane. (b) A summary of the bands found along the ΓX and ΓM cuts for different polarizations and their notation. Spectra excited with σ- and π-polarization allow the detection of initial state wavefunctions with in- and out-of-plane character, respectively. HB and EB denote hole and electron bands, respectively. The subscripts σ and π indicate the polarization which selects for the band. For example, HBσΓX corresponds to a hole band along ΓX observed with σ-polarization. The HBΓX and HBΓM bands are detected in both polarizations (see text).

| Polarization | Direction | Symmetry |
|---|---|---|
| π | Za (ΓX) | $(d_{xz} + d_{yz})$, $d_{xy}$, $d_{z^2}$ |
| σ | | $(d_{xz} - d_{yz})$, $d_{x^2-y^2}$ |
| π | Zb (ΓM) | $d_{xz}$, $d_{x^2-y^2}$, $d_{z^2}$ |
| σ | | $d_{xy}$, $d_{yz}$ |

**Table 1.** Orbital symmetry of bands detected in ARPES experiments for different photon polarizations and scanning directions. The electron detector is assumed to be located in the mirror plane. The allowed initial states, and thus the symmetry of the allowed states, are determined with the parity selection rules applied to the photoelectron matrix elements (see text).

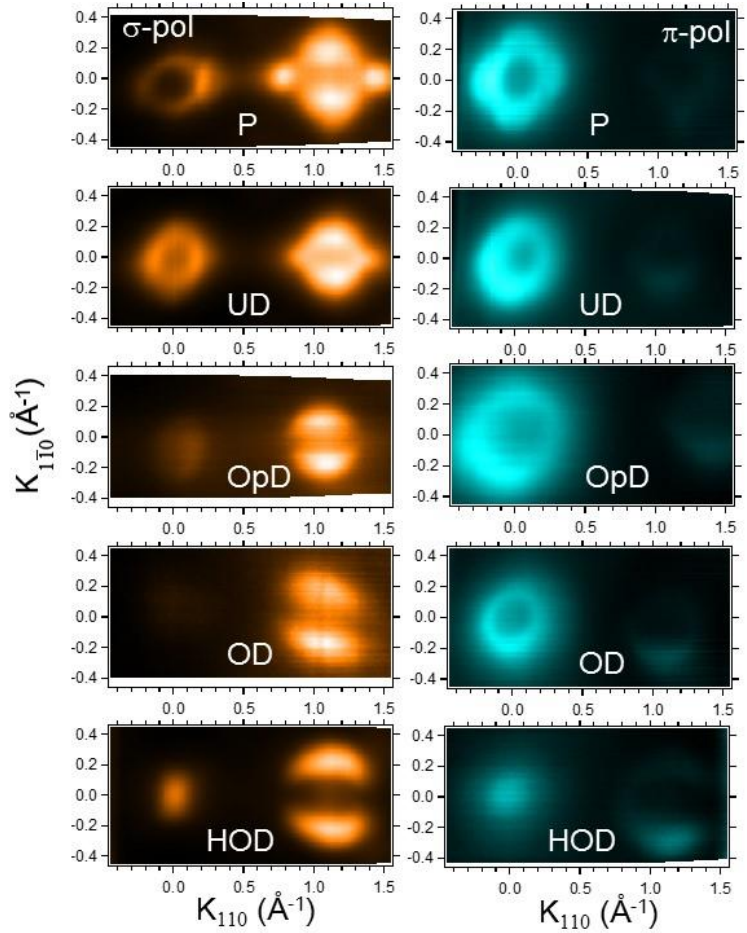

**Fig. 2. Fermi Surfaces.** Fermi Surfaces of Ba(Fe$_{1-x}$Co$_x$)$_2$As$_2$ for different Co concentrations, measured along the ΓM direction. The data were taken in the normal state at T = 28 K both with σ- and π-polarization (left and right panels, respectively) and with an incoming photon energy of 60 eV. The FS maps were generated by integrating the spectra over a window of 10 meV around E$_F$ for better visualization. Higher color brightness corresponds to higher intensity.

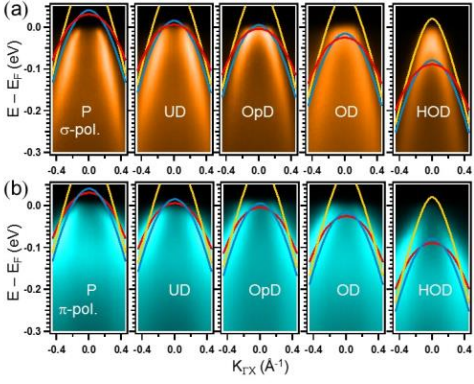
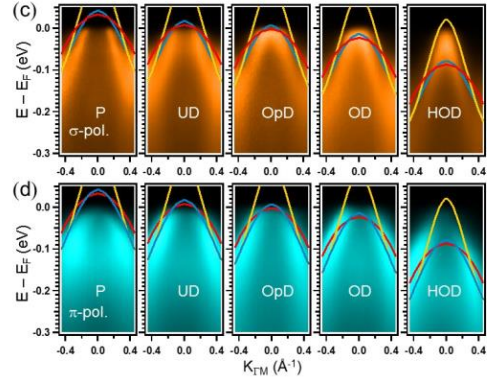
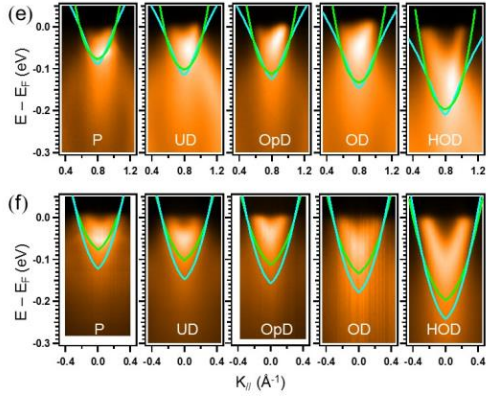

**Fig. 3. Image plots for data collected along the ΓX and ΓM direction, centered at the Γ (0,0) and M (π,π) points.** (a-d) Image plots of data collected at Γ (hole pocket) along the (a,b) ΓX and (c,d) ΓM direction with (a,c) σ- and (b,d) π-polarization. (e,f) Image plots of data collected at M (electron pocket) with σ-polarization along the (e) ΓX, and (f) ΓM direction. Data collected at M with π-polarization had extremely low intensity (cf. Fig. 2), and are not shown. The continuous lines denote the renormalized DFT bands (see text for details). Higher color brightness corresponds to higher intensity.

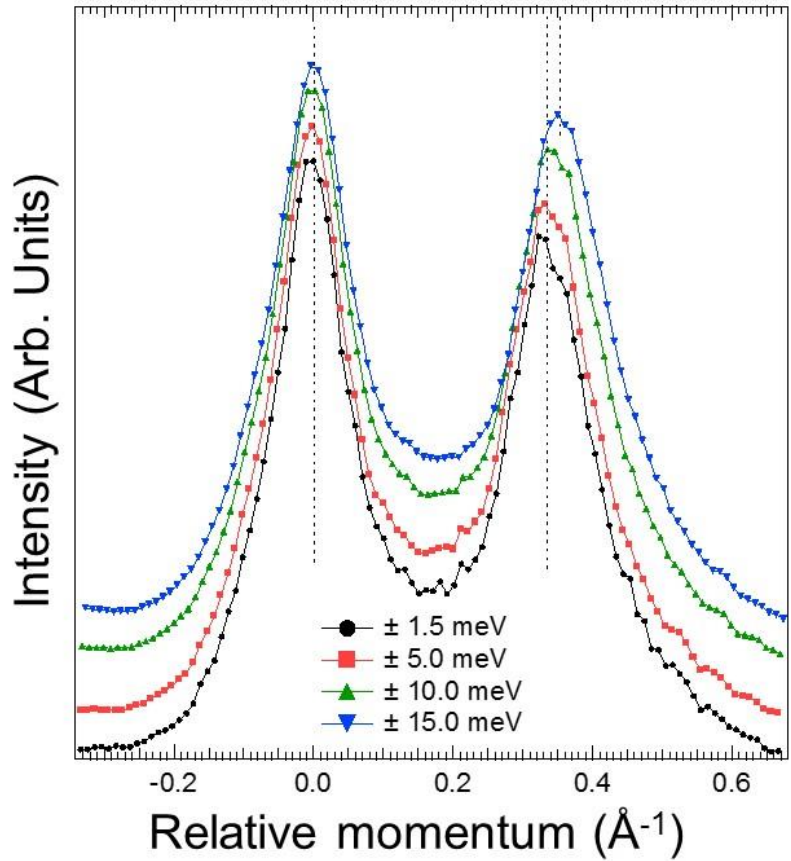

**Fig. 4. MDC spectra extracted using different integration windows at the Fermi level.** The data were collected from the parent compound at Γ (hole pocket) along the ΓM direction with σ-polarization. The MDC spectra were extracted using different integration windows at the Fermi level, as indicated in the legend. All of the peaks have been aligned on the left hand side of the plot in order to show clearly the systematic change of the positions of the peaks on the right hand side. The zero of the momentum scale has been set to coincide with the center of the peaks on the left hand side of the plot. Note also the substantial broadening introduced by the use of larger integration windows. The lines through the data points are guides to the eye.

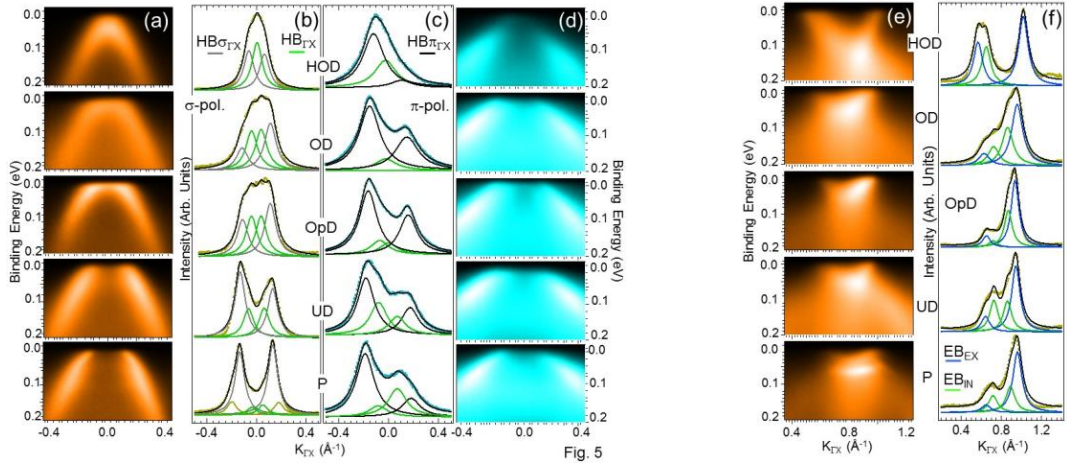

**Fig. 5. Image plots and MDC analysis for data collected along the ΓX direction, centered at the Γ (0,0) and M (π,π) points.** (a-d) Image plots, MDC spectra at $E_F$, and fits of data collected at Γ (hole pocket) with (a,b) σ- and (c,d) π-polarization. (e,f) Image plots, MDC spectra at $E_F$, and fits of data collected at M (electron pocket) with σ-polarization. Higher color brightness corresponds to higher intensity. A minimum number of peaks were used to fit the MDCs. The continuous lines through the data points are the results of the fits. At Γ (a-d), for both polarizations, two bands are always detected: one band visible only in σ polarization ($HBσ_{ΓX}$), one band visible only in π polarization ($HBπ_{ΓX}$), and one band visible ($HB_{ΓX}$) in both polarizations. For the parent compound there is one additional band visible in the σ-polarization. In the HOD compound the $HB_{ΓX}$ bands lie below $E_F$, and thus only the tails of the bands are visible, and appear as a single peak. (e,f) At M, two bands are detected, but they cannot be distinguished based on their orbital symmetry. They are labelled as an internal ($EB_{IN}$) and an external ($EB_{EX}$) band.

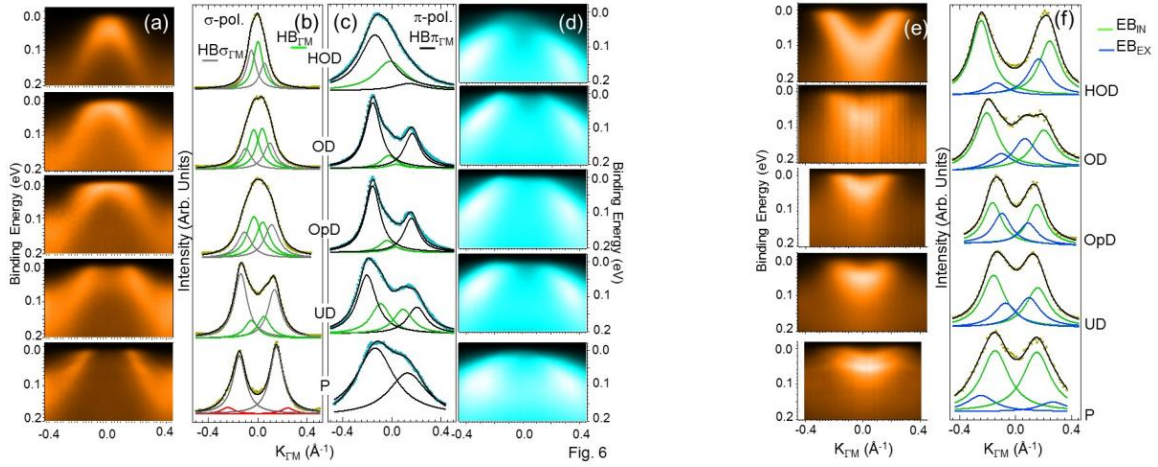

**Fig. 6. Image plots and MDC analysis for data collected along the ΓM direction centered at the Γ (0,0) and M (π,π) points.** (a-d) Image plots, MDC spectra at $E_F$, and fits of data collected at Γ (hole pocket) with (a,b) σ- and (c,d) π-polarization. (e,f) Image plots, MDC spectra at $E_F$, and fits of data collected at M (electron pocket) with σ-polarization. Higher color brightness corresponds to higher intensity. A minimum number of peaks were used in the fit of the MDCs, resulting in the continuous lines through the data. At Γ (a-d), for both polarizations, two bands are always detected: one band visible only in σ polarization (labeled $HB\sigma_{\Gamma M}$), one band visible only in π polarization (labeled $HB\pi_{\Gamma M}$), and one band visible in both polarizations (labeled $HB_{\Gamma M}$). For the parent compound there is one small additional band visible in the σ-polarization, which originates from the band folding induced by the structural phase transition T = 140 K. In the HOD compound the $HB_{\Gamma M}$ bands lie below $E_F$, and thus only the tails of the bands are visible, appearing as a single peak. (e,f) At the M point, two bands are detected, but they cannot be distinguished based on their orbital symmetry. They are labeled as the internal ($EB_{IN}$) and external ($EB_{EX}$) electron bands.

| Direction | Band | Symmetry |
|---|---|---|
| Za (ΓX) | HBπ$_{\Gamma X}$ | $d_z^2$ |
|  | HBσ$_{\Gamma X}$ | $d_{x^2-y^2}$ |
|  | HB$_{\Gamma X}$ | $d_{xz} \pm d_{yz}$ |
| Zb (ΓM) | HBπ$_{\Gamma M}$ | $d_{xz}, d_z^2$ |
|  | HBσ$_{\Gamma M}$ | $d_{xy}, d_{yz}$ |
|  | HB$_{\Gamma M}$ | $d_{xz}, d_{yz}$ |
| Zb (ΓM) | EB$_{IN}$, EB$_{EX}$ | $d_{xy}, d_{yz}$ |

**Table 2**. Orbital character of the bands detected with different photon polarization and scanning directions. The assignment of the orbital characters of the bands is discussed in the text.

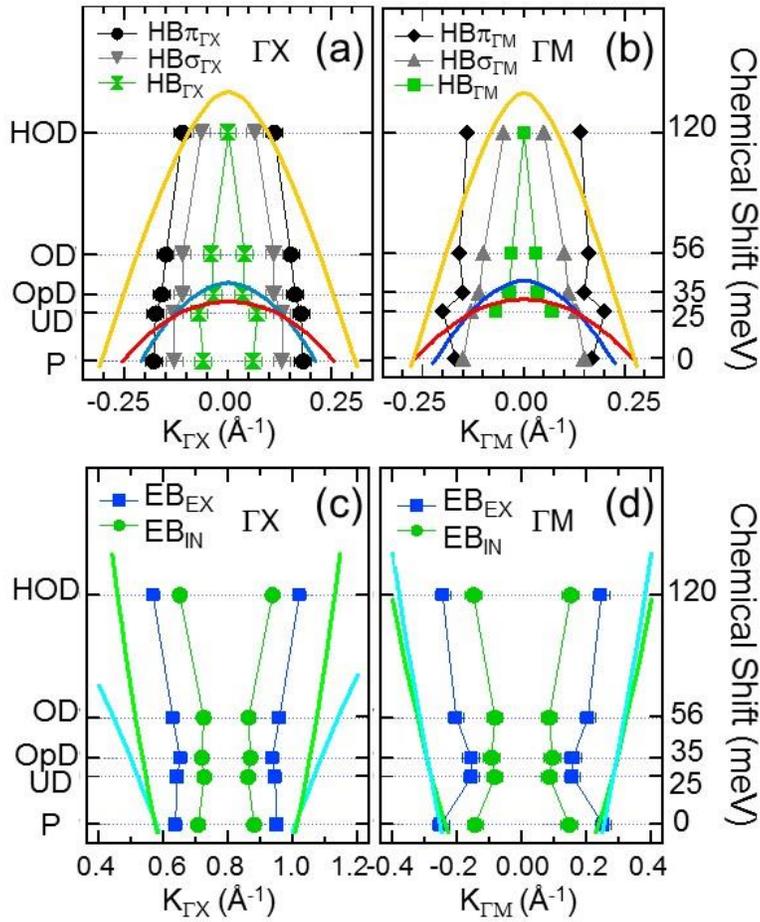

**Fig. 7. Co concentration dependence of the Fermi momenta and comparison to DFT results.** The Fermi momenta $k_F$, identified by the position of the peaks fitting the MDC shown in Figs. 5 and 6, are plotted for the different Co concentrations indicated on the left vertical scales. The right vertical scales indicate the corresponding shift of the chemical potential as calculated with DFT (see text). The lines through the data points are guides to the eye. The error bars on the Fermi momenta are comparable to the symbols size. (a)-(b) Result from the fitting of the MDC at the hole pocket along the (a) ΓX and (b) ΓM directions. (c)-(d) Result from the fitting of the MDC at the electron pocket along the (c) ΓX and (d) ΓM directions. The continuous lines are the DFT bands for the P compound. According to the DFT calculations, the band structure of the doped compounds is obtained by raising the chemical potential μ corresponding to the value of the chemical shift indicated on the right vertical scale in the figures.

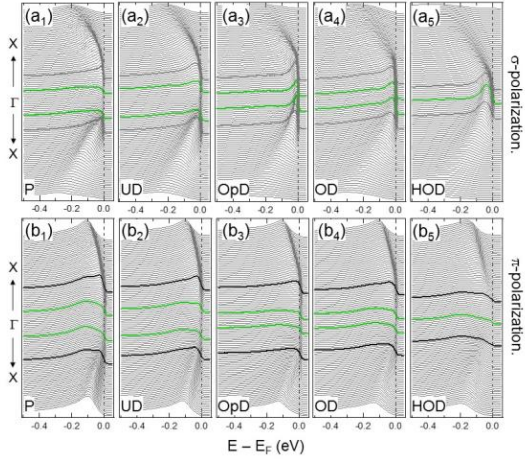
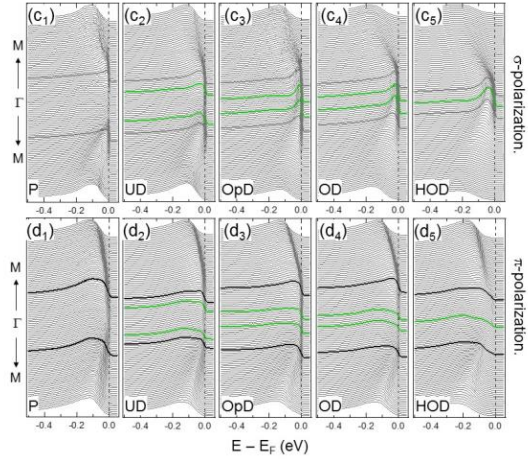
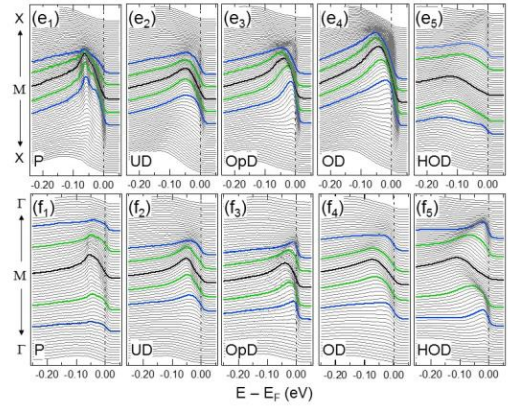

**Fig. 8. Energy Distribution Curves stacks.** Energy Distribution Curve (EDC) stacks, corresponding to the data shown in Figs. 5 and 6. Panels (a)-(d) show EDCs of data collected at the Γ point (hole pocket) along the (a),(b) ΓX and (c),(d) ΓM directions. The photon polarization is indicated in the figure. Panels (e)-(f) show EDCs of data collected in σ- polarization at the M point (electron pocket) along the (e) ΓX and (f) ΓM directions. The highlighted spectra denote the EDC at the Fermi momenta $k_F$. The outer EDC corresponds to the $k_F$'s of the (a) HBσΓX, (b) HBπΓX, (c) HBσΓM, (d) HBπΓM, and (e,f) $EB_{EX}$ bands. In (a)-(d), the inner EDC corresponds to $k_F$ for the (a,b) HBΓX and (c,d) HBΓM bands. In (e,f), the inner and outer EDCs correspond to the $k_F$'s for the $EB_{IN}$ and $EB_{EX}$ bands, respectively. The highlighted spectra at the center of the EDC stack denote the EDC taken at the M point (bottom of electron pocket).

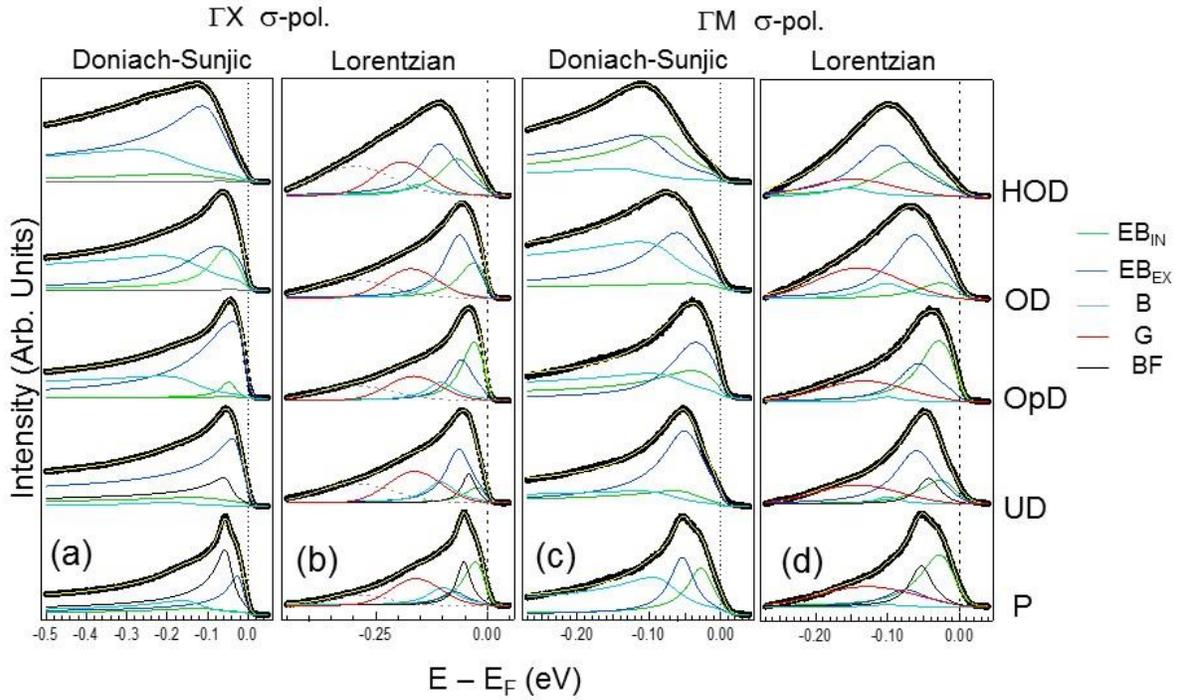

**Fig. 9. Co concentration dependence and fits of the EDCs at M.** EDCs extracted from the data at the M point (bottom of electron pocket) along the (a,b) $\Gamma X$ and (c,d) $\Gamma M$ directions in $\sigma$-polarization. In (a),(c) the EDC spectra are fit with the Doniach-Šunjić (DS) lineshape, while the spectra shown in (b),(d) are fit with Lorentzian (L) peaks and an additional Gaussian peak denoted as G, after subtraction of a Shirley background. Peak B denotes a broad hole-like band that never crosses $E_F$, but intersects the $EB_{IN}$ and $EB_{EX}$ bands close to the M point. The narrow peak (BF) visible in the spectra along the $\Gamma X$ direction in the P and UD compounds accounts for the band folding due to the orthorhombic distortion. In panel (b), the dashed line denotes an additional background necessary to fit the longer tail of the spectra. The use of the DS lineshape does not produce satisfactory fits. For clarity, the spectra in the figures were normalized to the same height, so overall intensities cannot be directly compared.

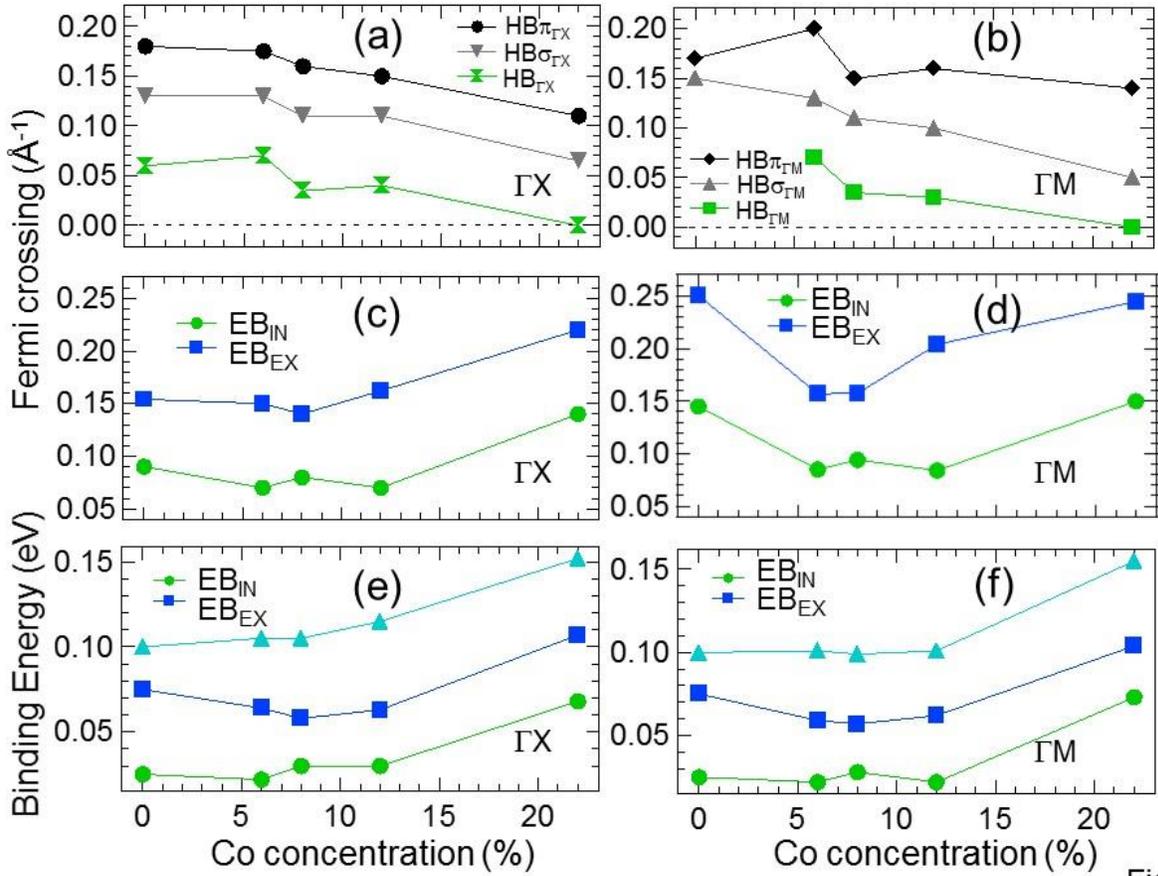

**Fig. 10. Co concentration dependence of the Fermi momenta and Binding Energy of the $EB_{IN}$ and $EB_{EX}$ bands obtained from fitting the EDCs at the M point.** The Fermi momenta, identified by the position of the peaks in the MDCs fits, are plotted as a function of Co concentration. The error bars (± 0.005 Å$^{-1}$) are smaller than the size of the symbols. The continuous lines through the data points are a guide to the eye. (a,b) Result from the fitting of the MDCs at the hole pocket along the (a) ΓX and (b) ΓM directions. (c,d) Result from the fitting of the MDCs at the electron pocket along the (c) ΓX and (d) ΓM directions. (e,f) Doping dependence of the binding energy of the $EB_{IN}$ and $EB_{EX}$ bands obtained from fitting the EDCs at the M point (bottom of the electron pocket). The error bars are comparable to the size of the symbols. For lower Co concentrations, the bottom of the $EB_{EX}$ band shifts toward $E_F$, while the bottom of the $EB_{IN}$ band exhibits an irregular, non-monotonic shift. For different Co concentrations, only in the HOD sample does the bottom of the $EB_{IN}$ and $EB_{EX}$ bands shifts to higher BE. This behavior is clearly at odds with a rigid shift of the band structure, since in this case the bottom of the bands would drift progressively away from $E_F$. Also shown, denoted with the triangle symbol, is the position of peak B in Fig. 9, i.e. the maximum of the band that never crosses $E_F$ (see text).

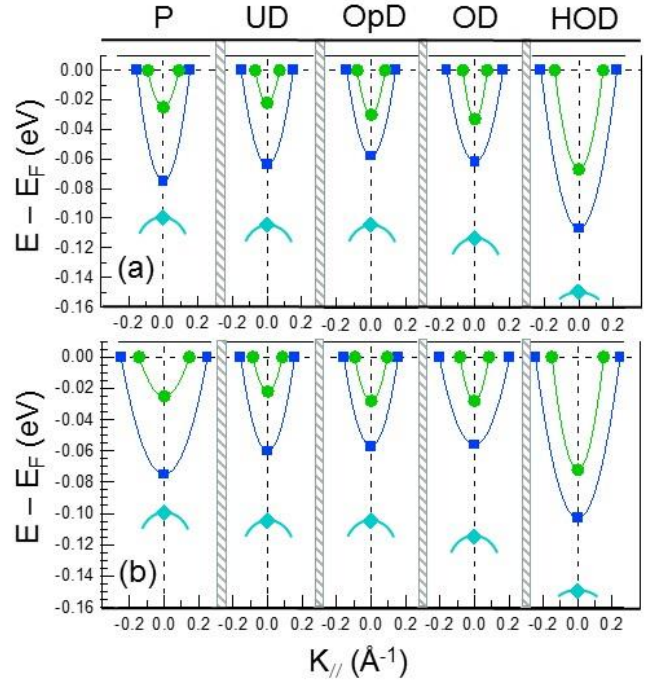

Fig. 11

**Fig. 11. Schematic illustration of the Co concentration dependence of the electronic structure at the M point (electron pocket).** The positions of the Fermi momenta from the MDC analysis (cf. Figs. 10(c),(d)) and the BE of the bottom of the electron pocket obtained from the fitting of the EDC spectra (cf. Figs. 10 (c),(d)) for the (a) ΓX and (b) ΓM directions are plotted for different Co concentration. Also shown, denoted with the diamond symbol, is the position of the the maximum of the band that never crosses $E_F$ (cf. peak B in Fig. 9, Figs. 10 e) and f)). The continuous lines are guides for the eye.

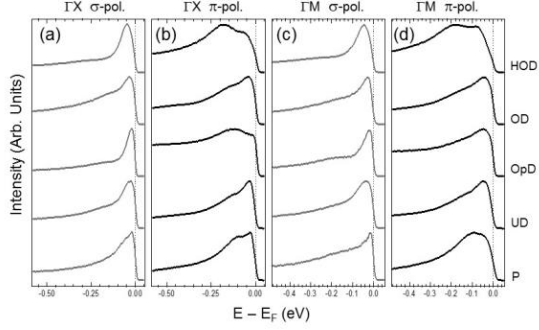 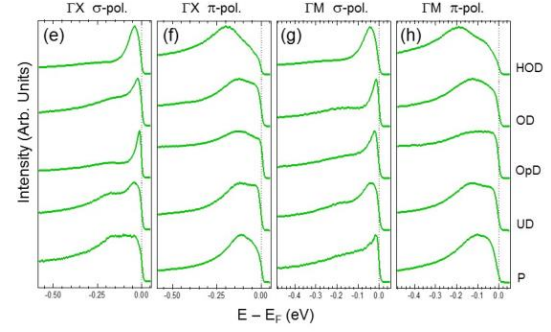

Fig. 12

**Fig. 12. Co concentration dependence of the EDCs at the Fermi momenta ($k_F$) for the bands at $\Gamma$ (hole pocket).** (a)–(d) Outer band EDCs extracted from the image plots corresponding to $k_F$ for cuts along the (a)-(b) $\Gamma X$ and (c)-(d) $\Gamma M$ directions. (e)-(h) Inner band EDCs extracted from the image plots corresponding to $k_F$ for cuts along the (e)-(f) $\Gamma X$ and (g)-(h) $\Gamma M$ directions. The photon polarization is indicated on top of the panels. For clarity, the spectra in the figures were normalized to the same height, so a comparison of the intensities is meaningless.

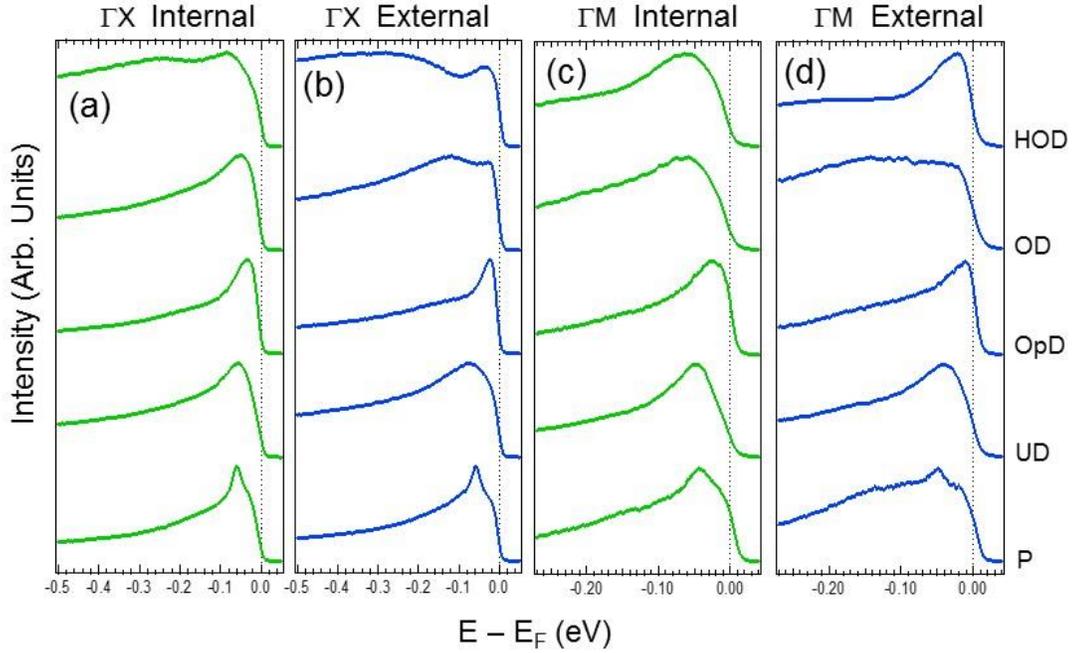

Fig. 13

**Fig. 13. Co concentration dependence of the EDCs at the Fermi momenta ($k_F$) for the bands at M (electron pocket).** EDCs extracted from the image plots corresponding to $k_F$ for cuts along the (a)-(b) ΓX and (c)-(d) ΓM directions. All of the spectra have been taken with σ-polarization. The spectra sharpen in correspondence of optimal doping. For clarity, the spectra in the figures were normalized to the same height, so a comparison of the intensities is meaningless.

## APPENDIX 1

**Density functional theory (DFT) calculations** - The band structure calculations were carried out with the PBE-GGA functional in the virtual crystal approximation scheme. For the parent compound BaFe$_2$As$_2$ the experimental crystal structure at 175 K was used, as described by Q, Huang et al. [19], in which the As height with respect to the Fe plane differs from the relaxed optimized structure provided by Local Density Approximation (LDA). The calculations were done using the experimental height. The results of the band structure calculations highlighting orbital characters of the bands are shown in Fig. S1.

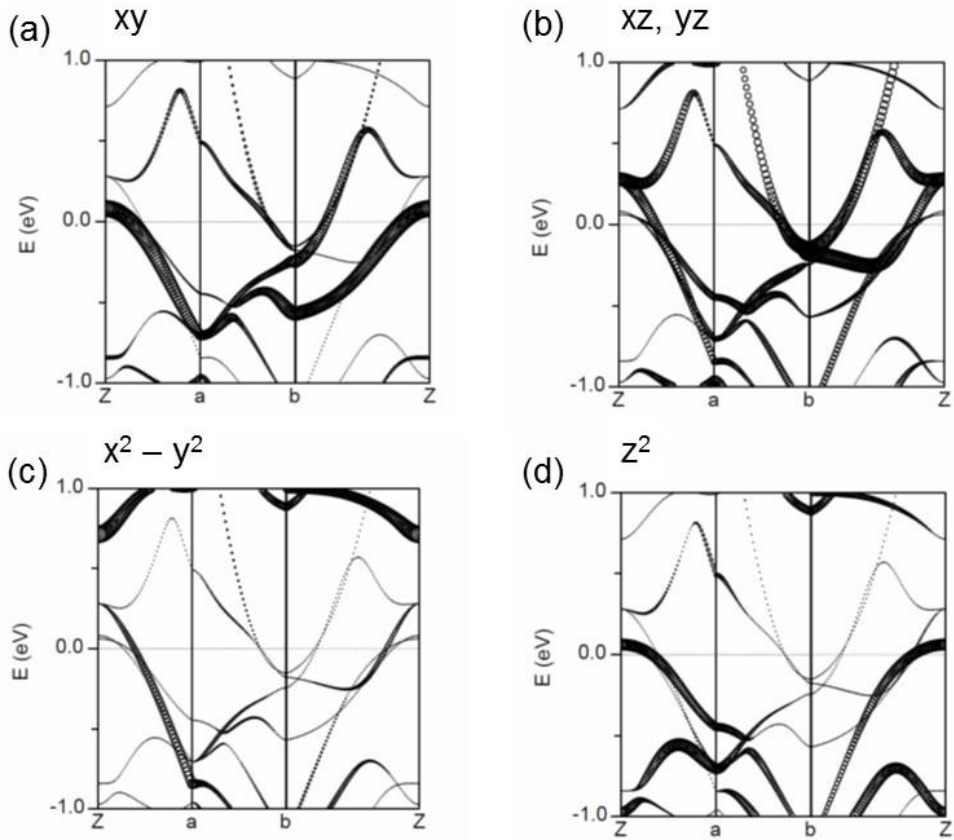

**Fig. S1**. Results of the DFT calculations in the parent compound. The calculations are pertinent to the plane containing the Z point, i.e. the zone boundary along the vertical $k_z$ direction. As explained in the text, we refer to the high symmetry directions Za and Zb as ΓX and ΓM, respectively. In each panel the size of the markers indicate the amount of orbital character contributed to each band. (a) xy, (b) xz, yz, (c) $x^2 - y^2$, (d) $z^2$.